# Investigation of Level Statistics by Generalized Brody Distribution and Maximum Likelihood Estimation Method

M. A. Jafarizadeh[a,b1], N. Fouladi[c], H. Sabri[c2], B. R. Maleki[c]

[a] Department of Theoretical Physics and Astrophysics, University of Tabriz, Tabriz 51664, Iran.
[b] Research Institute for Fundamental Sciences, Tabriz 51664, Iran.
[c] Department of Nuclear Physics, University of Tabriz, Tabriz 51664, Iran.

---

[1] E-mail: jafarizadeh@tabrizu.ac.ir
[2] E-mail: h-sabri@tabrizu.ac.ir

**Abstract**

With generalizing the Brody distribution to include the Poisson, GOE and GUE limits and with employing the maximum likelihood estimation technique, the spectral statistics of different sequences were considered in the nearest neighbor spacing statistics framework. The ML-based estimated values for the parameters of generalized distribution propose more precisions in compare to the predictions of other distributions. The transition in the level spacing statistics of different systems were described by the distances of ML-based predictions for generalized distribution to three limits which determined by KLD measures.

**Keywords:** Nearest Neighbor Spacing Distribution (NNSD); Maximum Likelihood Estimation (MLE); Brody distribution; Gaussian Orthogonal Ensemble (GOE); Gaussian Unitary Ensemble (GUE). Kullback-Leibler Divergence (KLD).

**PACS:** 05.45.Mt; 24.60.-k; 02.50.Tt

**Introduction**

The Random Matrix Theory (RMT) is known as a main tool in describing the statistical distribution of the energy-eigenvalues for the quantum counterpart of a classical chaotic system. It has been supposed that eigenvalues belonging to different irreducible representations of the symmetry group are statistically independent and obey Gaussian Orthogonal Ensemble (GOE) or Gaussian Unitary Ensemble (GUE) spectral statistics [1-2]. The GUE statistics usually is expected for spectra of non-time-reversal invariant systems. Recently it has been exposed [3-5], time-reversal invariant systems with discrete symmetries may display, in certain irreducible subspaces, the spectral statistics corresponding to the Gaussian-unitary ensemble (GUE) rather than to the expected orthogonal one (GOE). Different statistics [6-8] have been proposed to exhibit the statistical situation of systems in related to regular (Poisson limit) and chaotic (GOE or GUE) limits while the Nearest Neighbor Spacing Distribution (NNSD) statistics is the observable most commonly used to analyze the short-range fluctuation properties in the considered spectra.

In the common descriptions, a comparison of NNSDs with well known distributions such as Brody and *etc* [9-11] has been carried by Least Square Fit (LSF) technique [12-13]. The estimated value(s) for distribution's parameter(s) explores the statistical situation of considered systems [14-15]. These distributions describe interpolations between only GOE and Poisson limits by one parameter and therefore one had to use different statistics such as $\Delta_3(L)$ to consider the statistical properties which closer to GUE limit. Also, the LSF technique proposes some unusual uncertainty for estimated values and also a deviation to more chaotic statistics for considered systems [16].
In the present paper, to describe the spectral statistics which are closer to GUE in the NNSD framework, the Brody distribution is generalized to include GOE, GUE and Poisson limits. Also, with more parameters in this generalized distribution in compare to other ones, one can expect more precisions in the estimation processes. Also to overcome the disadvantages of LSF-based estimation procedures, the Maximum Likelihood Estimation (MLE) technique was employed which suggest more regularity and also high precisions for estimated parameters (estimated values yield accuracies which are closer to Cramer-Rao Lower Bound (CRLB))[16].

To consider the advantages of the generalized Brody distribution and also MLE technique in spectral investigations, we prepared some sequences. With employing all the empirical available data [17-18] for even mass nuclei introduced in Ref.[8], three sequences constructed, i.e. all considered even mass nuclei, spherical and deformed even-mass nuclei. Sequences constructed of $2^+$ and $4^+$ levels of nuclei in which the spin-parity $J^\pi$ assignment of at least five consecutive levels are definite. Also with employing the sequences introduced in Ref.[19], the transition between GUE and GOE limits for electrons in graphene was considered.

The ML-based estimated values for the generalized Brody distribution propose more precisions in compare to Brody one in same sequences. Also, the determined CRLBs, offer the least bound for generalized distribution in these sequences. The distances of ML-based predictions for generalized distribution to both GOE and GUE limits reveal the results of Huang *et al* in Ref.[19], namely a transition of GUE to GOE in the spectral statistics of electrons in graphene due to a weak magnetic field.

This paper is organized as follows: section 2 briefly summarizes details about statistical investigation which includes unfolding procedure, generalized Brody distribution and also MLE technique which applied to the new distribution. In section 3, numerical results are presented while Section 4 is devoted to summarize and some conclusion based on the results given in section 3. The paper ends with appendices containing the details and related calculations of generalized Brody distribution and CRLB.

## 2. Statistical investigation

The spectral fluctuations of low-lying nuclear levels have been considered by different statistics such as Nearest Neighbor Spacing Distribution (NNSD) [1-2], Dyson-Mehta $\Delta_3(L)$ statistics [7] and *etc* which based on the comparison of statistical properties of nuclear spectra with the predictions of Random Matrix Theory (RMT). The NNSD, or $P(s)$ functions, is the observable most commonly used to analyze the short-range fluctuation properties in the nuclear spectra. To compare the different sequences to each other, each set of energy levels must be converted to a set of normalized spacing, namely, each sequence must be unfolded. To unfold our spectrum, we have to use some levels with same symmetry. This requirement is equivalent with the use of levels with same total quantum number ($J$) and same parity. For a given spectrum $\{E_i\}$, it is necessary to separate it into the fluctuation part and the smoothed average part, whose behavior is nonuniversal and cannot be described by RMT [1-2]. To do so, we count the number of the levels below $E$ and write it as

$$N(E) = N_{av}(E) + N_{fluct}(E) \qquad , \qquad (2.1)$$

Then we fix the $N_{av}(E_i)$ semiclasically by taking a smooth polynomial function of degree 6 to fit the staircase function $N(E)$. We obtain finally, the unfolded spectrum with the mapping

$$\{\tilde{E}_i\} = N(E_i) \quad , \tag{2.2}$$

This unfolded level sequence $\{\tilde{E}_i\}$ is obviously dimensionless and has a constant average spacing of 1 but the actual spacing exhibits frequently strong fluctuation. The nearest neighbor level spacing is defined as $s_i = (\tilde{E}_{i+1}) - (\tilde{E}_i)$. Distribution $P(s)$ will be in such a way in which $P(s)ds$ is the probability for the $s_i$ to lie within the infinitesimal interval $[s, s+ds]$. For nuclear systems with time reversal symmetry which spectral spacing follows Gaussian Orthogonal Ensemble (GOE) statistics, the NNS probability distribution function is well approximated by Wigner distribution [1-2]

$$P(s) = \frac{1}{2}\pi s e^{-\frac{\pi s^2}{4}} \quad , \tag{2.3}$$

This exhibits the chaotic properties of considered spectra. On the other hand, the NNSD of systems with regular dynamics is generically represented by Poisson distribution

$$P(s) = e^{-s} \quad , \tag{2.4}$$

Different statistical investigations accomplished on the nuclear system's spectra, propose intermediate situations between these limits. To compare the spectral statistics with regular and chaotic limits and also exhibit interpolation between them, different distribution functions have been proposed [9-11]. One of the popular distributions is Brody distribution [9]

$$P(s) = b(1+q)s^q e^{-bs^{q+1}} \quad , \quad b = [\Gamma(\frac{2+q}{1+q})]^{q+1} \tag{2.5}$$

Considers a power-law level repulsion and interpolates between the Poisson ($q=0$) and Wigner ($q=1$) limits. All well-known distributions such as Brody, Berry-Robnuk [10] and Abul-Magd [11] describe only the Poisson and GOE limits and interpolation between them and therefore, one had to use the $\Delta_3(L)$ statistics to describe the spectral statistics which closer to GUE limit, similar to procedures have done in Refs.[19-24]. In the following, we generalized the Brody distribution (which has the most efficiency in compare to other distributions [16]) to consider the statistical properties of different systems in compare to all Poisson, GOE and GUE limits in the NNSD-based statistics.

## 2.1. Generalized Brody distribution

The phenomenon of level repulsion in energy spectra has been investigated in different papers [14-16]. Different distribution functions have been suggested to describe the statistical behavior of considered system between Poisson (order limit) and GOE limit of Random Matrix Theory (RMT) [1-2]. In order to investigate the spectral statistics in compare to all Poisson (order), GOE (Wigner or chaotic) and GUE limits and also estimate with more precision, we generalized Brody distribution which derived from Wigner surmise.

The nearest neighbor spacing of Gaussian orthogonal ensemble was distributed as Eq.(2.3). On the other hand, the nearest neighbor spacing of Gaussian unitary ensemble can be described by [1-2]

$$P(s) = \frac{32}{\pi^2} s^2 e^{-\frac{4s^2}{\pi}} \quad , \tag{2.6}$$

We extended both (2.3) and (2.6) relations by means of ansatz

$$P(s) = b(1+q)(\alpha s^q + \beta s^{q+1}) e^{-bs^{q+1}} \quad , \tag{2.7}$$

Which interpolate between Poisson $(q=0 \ \& \ \beta=0)$, GOE or Wigner limit $(q=1 \ \& \ \beta=0)$ and GUE limit $(q=1 \ \& \ \alpha=0)$. Since the $P(s)$ must be normalized;

$$\int_0^\infty P(s)\,ds = 1 \qquad \& \qquad \int_0^\infty s\,P(s)\,ds = 1$$

We can obtain the constants of (2.7) as

$$\alpha = 1 - \frac{(\frac{\Gamma[\frac{q+2}{q+1}]}{b^{\frac{1}{q+1}}})^2 - \frac{\Gamma[\frac{q+2}{q+1}]}{b^{\frac{1}{q+1}}}}{(\frac{\Gamma[\frac{q+2}{q+1}]}{b^{\frac{1}{q+1}}})^2 - \frac{\Gamma[\frac{q+3}{q+1}]}{b^{\frac{2}{q+1}}}} \qquad , \qquad \beta = \frac{(\frac{\Gamma[\frac{q+2}{q+1}]}{b^{\frac{1}{q+1}}}) - 1}{(\frac{\Gamma[\frac{q+2}{q+1}]}{b^{\frac{1}{q+1}}})^2 - \frac{\Gamma[\frac{q+3}{q+1}]}{b^{\frac{2}{q+1}}}} \qquad (2.8)$$

In the following, we would employ the MLE technique to determine the estimators of generalized Brody distribution's parameters.

### 2.2. The ML- based relations for generalized Brody distribution

In common considerations [6-8], one can concern a least square fit (LSF) of Brody distribution to considered sequences while the value of distribution's parameter characterizes chaotic (Wigner limit) or regular (Poisson limit) dynamics. As have described in Refs.[12-13,16], due to the high level variances of estimators, the LSF-based estimated values have some unusual uncertainties. On the other hand, the LSF technique is on firm theoretical grounds when it can reasonably be assumed, the deviations of the observations from the expectations of the true theory are independently, identically and normally distributed, therefore, one can expect a deviation to chaotic dynamics by prediction of LSF method [16]. Consequently, it is almost impossible to do any reliable statistical analysis in some sequences. Recently, we have employed the Maximum Likelihood Estimation (MLE) technique [12-13] to estimate every distribution's parameter which provides more precisions, i.e. low uncertainties. It means, the estimated values yield accuracies which are closer to CRLB. The MLE estimation procedure has been described in detail in Ref [16]. Here, we outline the basic ansatz and summarize the results. At first, we must generate the appropriate likelihood functions to estimate $\alpha, \beta, q$ and $b$. Due to some problems which concern in the maximizing the likelihood function contains Gamma functions, we consider $\alpha$ and $\beta$ as independent quantities and define estimators for them but as would present in the following (Appendix I), this assumption wouldn't effect on overall definition of these quantities. To estimate, we try to use the products of the generalized Brody distribution functions as a likelihood function, namely

$$L(q,b,\alpha,\beta) = \prod_{i=1}^n b(1+q)(\alpha s_i^q + \beta s_i^{q+1})e^{-bs_i^{q+1}} = (b(1+q))^n \prod_{i=1}^n (\alpha s_i^q + \beta s_i^{q+1})e^{-bs_i^{q+1}} \qquad , \qquad (2.9a)$$

With using the fact that $L(q,b,\alpha,\beta)$ and $\ln(L(q,b,\alpha,\beta))$ have maximum value for the same values of quantities, we use

$$\ln L(q,b,\alpha,\beta) = n\ln(b(1+q)) + \sum_{i=1}^n \ln(\alpha s_i^q + \beta s_i^{q+1}) - b\sum_{i=1}^n s_i^{q+1} \qquad , \qquad (2.9b)$$

Then, taking the derivative of the log of likelihood function (2.9b) respect to its parameters and set them to zero, i.e., maximizing likelihood functions, the following relations for desired estimators (see Appendix (I) for more details) is obtained

$$\frac{\partial \ln L(q,b,\alpha,\beta)}{\partial q} = 0 \quad \Rightarrow \quad f_1 : \frac{n}{1+q} + \sum_{i=1}^{n} \ln s_i - b \sum_{i=1}^{n} \ln s_i \, s_i^{q+1} \quad \text{for } q \quad (2.10a)$$

$$\frac{\partial \ln L(q,b,\alpha,\beta)}{\partial b} = 0 \quad \Rightarrow \quad f_2 : \frac{n}{b} - \sum_{i=1}^{n} s_i^{q+1} \quad \text{for } b \quad (2.10b)$$

$$\frac{\partial \ln L(q,b,\alpha,\beta)}{\partial \alpha} = 0 \quad \Rightarrow \quad f_3 : \sum_{i=1}^{n} \frac{s_i^q}{\alpha s_i^q + \beta s_i^{q+1}} \quad \text{for } \alpha \quad (2.10c)$$

$$\frac{\partial \ln L(q,b,\alpha,\beta)}{\partial \beta} = 0 \quad \Rightarrow \quad f_4 : \sum_{i=1}^{n} \frac{s_i^{q+1}}{\alpha s_i^q + \beta s_i^{q+1}} \quad \text{for } \beta \quad (2.10d)$$

Now, the parameters $\alpha, \beta, q$ and $b$ can estimate by high precision via solving the above equations by Newton-Raphson iteration method as have explained in Appendix (I). To estimate by MLE technique which provide more precision, we have followed the prescription explained in Ref.[16], namely the ML-based estimated parameters correspond to the converging values of iterations relations (I-10 to13) of Appendix (I), where as an initial values we have chosen the values of parameters obtained by LSF method.

Also, we have employed the Cramer-Rao Lower Bound (CRLB) inequality to describe the variances of considered estimators in the iteration procedures [12-13,16]. Namely, the value of this quantity for different distributions in the similar sequences characterizes the efficiency of distribution, i.e. the lowest CRLB suggest distribution as the best one in the statistical analyses. The CRLB for vector functions of vector parameters defined as [13];

$$\text{CRLB}: \frac{\partial \rho(\theta)}{\partial \theta^T} [F(\theta)]^{-1} \frac{\partial \rho^T(\theta)}{\partial \theta} \quad , \quad \theta \equiv \alpha, \beta, b \text{ and } q \quad (2.11)$$

where $F(\theta)$ and $\rho(\theta)$ represent Fisher information and the estimator of considered quantity (see Appendix (II) of for more details).

As have explained in the previous subsection, one had to use the $\Delta_3$ statistics for describing the transition between regular (Poisson) and one of chaotic (GOE or GUE) limits. The closer approaches to one of these limits (which have been realized qualitatively from their curves) are considered as the measure in this method while similar comparison wouldn't possible in the NNSD-based statistics. The predictions of generalized Brody distribution and their distances to all limits (which determined by KLD measurers) provide a quantitatively description in the similar systems. The Kullback-Leibler Divergence (KLD) measure was defined as [12-13];

$$D_{KL}(P \parallel Q) = \sum_{s_i} P(s_i) \, \log \frac{P(s_i)}{Q(s_i)} \quad , \quad (2.12)$$

In which it would display closer distances between two distributions if $D_{KL}(P \parallel Q) \to 0$ while we would consider the ML-based estimated generalized Brody distribution as $P(s_i)$ and GOE or GUE distributions as $Q(s_i)$ in our analyses. To analyze these situations, we take the following process. At first, by Eq.(2.12), we calculate the distances of ML-based estimated generalized Brody distribution to all Poisson, GOE and GUE limits. The smaller value between these distances explores the spectral statistics.

## 3. Numerical results of generalized Brody distribution in spectral analyses

In the estimation processes, one would expect more accuracy for distributions which have more parameters in compare to other ones. To compare the precisions of the Brody (which provide more accuracy in compare to other distributions [16]) and generalized Brody distributions, we consider the spectral statistics of some sequences. To prepare sequences by the available empirical data [17-18], we have followed the same method given in Ref.[8]. We consider nuclei in which the spin-parity $J^{\pi}$ assignments of at least five consecutive levels are definite. In cases where the spin-parity assignments are uncertain and where the most probable value appeared in brackets, we admit this value. We terminate the sequence in each nucleus when we reach at a level with unassigned $J^{\pi}$. We focus on the $2^+$ and $4^+$ levels of even mass nuclei for their relative abundances. With using nuclei have been introduced in Table (1) of Ref.[8] and unfolding procedure, three sequences prepared, namely, sequence included all considered even-mass nuclei, sequences contain deformed and spherical even-mass nuclei. The ML-based estimated values of Brody and generalized Brody distribution's parameters presented in the Table 1. The KLD measures which describe the distances of ML-based predictions for these distributions to the Poisson limit listed in Table1, too.

Table1. The ML-based estimated values for Brody and Generalized Brody distributions. The KLD measures suggest similar statistics in different sequences.

| Statistical criterions | All even mass nuclei | Spherical even mass nuclei | Deformed even nuclei |
|---|---|---|---|
| "q" Brody distribution's parameter | $0.18 \pm 0.02$ | $0.28 \pm 0.04$ | $0.12 \pm 0.03$ |
| $D_{KL}$(ML-based Brody $\parallel$ Poisson) | 0.852 | 1.361 | 0.244 |
| "q" gene.Brody distribution's parameter | $0.15 \pm 0.009$ | $0.23 \pm 0.02$ | $0.10 \pm 0.01$ |
| $D_{KL}$(ML-based gene.Brody $\parallel$ Poisson) | 0.690 | 1.119 | 0.193 |

The ML-based quantities and also the KLD measures for generalized Brody distribution propose similar statistics in different sequences, namely, deformed nuclei describe more regular dynamics in compare to other sequences where an obvious reduction in uncertainties are apparent in compare to the results of Brody distribution.

The KLD measures which describe the distances of ML-based predictions for generalized Brody distribution to Poisson limit propose a more deviation to regular dynamics, even more than the predictions of Brody distribution, i.e. the smaller $D_{KL}$ measures in the same sequences.

Also, the smaller uncertainties and therefore smaller CRLBs for generalized Brody distribution suggest it as distribution with the most efficiency in statistical investigations.

On the other hand, to consider the transition of spectral statistics between different limits of RMT, we have employed the sequences introduced in Ref.[19], namely the spectral statistics of electrons in graphene billiards due to the variations of magnetic fields. In the absence of magnetic field, electrons in graphene around the Dirac point obey the same massless particles equation in free space while describe the level statistics closer to GOE statistics. With considering a magnetic field on graphene, the true time reversal symmetry is broken and consequently, a transition occurs in the level statistics from GOE to GUE statistics. Also, if the magnetic field is sufficiently strong, around the Dirac point where the density of states is low, the energy levels are quantized into Landau levels and exhibit deviation from GUE to GOE limits. The considered sequences have been described in detail in Ref.[19]. Here, we briefly outline the basic ansatz and summarize the results.

With employing the tight-binding Hamiltonian as $\hat{H} = \sum (-t_{ij})|i\rangle\langle j|$, where the summation is over all pairs of nearest neighboring atoms and $t_{ij} = t \exp[-i \frac{2\pi}{\varphi_0} \int_{r_j}^{r_i} dr.A]$ is the nearest-neighbor hopping energy. Also, $A = (-By, 0, 0)$ is the magnetic vector potential for a perpendicular uniform magnetic field, $\phi_0 = h/e = 4.136 \times 10^{-15} \, Tm^2$ is the magnetic flux quanta and $t \approx 2.8 \, ev$ [19-20]. We would consider three cases in our description, $\phi = 0$, namely no magnetic field, $\phi = \phi_0 / 8000 \, (\sim 10T)$ for weak magnetic field and $\phi = \phi_0 / 800 \, (\sim 100T)$ for strong magnetic field. Though, graphene confinements have the geometric shape of chaotic billiards, we consider the Africa Billiard shape in our investigation [19]. For a relativistic spin-half particle, Berry *et al* obtained the smoothed spectral staircase function for positive eigenvalues as $\langle N(k) \rangle = Ak^2/4\pi - 1/12$ [20,24]. For our considered graphene billiard around the Dirac point, we have $E = \hbar v_F k$ where $v_F = \sqrt{3}ta/2\hbar$. With employing the linear-momentum relation for graphene, the smoothed counting staircase function with respect energy would be $\langle N(E) \rangle = \alpha E^2$ ($\alpha = A/2\pi\hbar^2 v_F^2$) which is known as unfolding spectra[22-24]. In the following, we would consider $x_n \equiv \langle N(E_n) \rangle$ as the unfolded spectra and $S_n = x_{n+1} - x_n$ regards as the nearest neighbor spacing. With using these sequences, the spectral statistics of graphene investigated in the three considered cases of magnetic fields. To measure the distances of ML-based generalized Brody distribution from the Wigner surmise, we have used the following quantity which similar to KLD measures describe distances between distributions [25];

$$\Delta = \frac{\sum_i [F^{GOE}(s_i) - F(s_i)]^2}{\sum_i [F^{GOE}(s_i) - F^{GUE}(s_i)]^2}$$

Where $F^{GOE}$, $F^{GUE}$ and $F$ are accumulated distributions derived from $P^{GOE}$, $P^{GUE}$ and ML-based estimated generalized Brody distribution respectively. Consequently, one can expect a closer approach to GOE limit if $\Delta = 0$ while $\Delta = 1$ represent a deviation to GUE limit. Table 2 display the $\Delta$ values while the NNSD figures for considered three cases of magnetic fields are presented in Figure1.

Table2. The relative distances of ML-based predictions for generalized Brody distribution to both GOE and GUE limits, $\Delta$, for three cases of magnetic fields which applied to electrons in graphene.

| Sequence | electron in graphene without magnetic field $\phi_0 = 0$ | electron in graphene with weak magnetic field $\phi = \phi_0/8000$ | electron in graphene with strong magnetic field $\phi = \phi_0/800$ |
|---|---|---|---|
| $\Delta$ | 0.09 | 0.91 | 0.06 |

As have proposed in the Ref.[19], the distances of ML-based predictions for the new distribution suggest a deviation to GUE limit for electrons in graphene if we add a weak magnetic field, i.e. the broken time reversal symmetry due to a magnetic field cause to this deviation in level statistics. On the other hand, in the absence of magnetic field and also while the magnetic field is sufficiently strong, the evaluated distances suggest a closer approach to GOE statistics.

From these tables and figures, we see the apparent reductions of the uncertainties for the ML-based estimated values for generalized Brody distribution in compare to Brody distribution. Also, with employing the KLD measures to exhibit distances of estimated function to both GOE and GUE limits, one would consider the transition of spectral statistics between different limits of RMT in the NNSD statistics framework with high precisions.

## Summary

In summary, we generalized the Brody distribution to consider the spectral statistics in general case with high accuracy. With employing the MLE technique, the required estimators prepared which estimate the parameters of distribution with more precision. In some sequences prepared by all the available empirical data, the ML-based estimated values and also CRLBs, suggest reductions of uncertainties. Also with using the KLD measures, one would describe the transition of spectral statistics of considered systems between different limits of RMT in the nearest neighbor spacing statistics framework. These results may be proposed this generalized distribution for spectral investigations of systems with chaotic dynamics between unitary and orthogonal limits.

# Appendix

## Appendix I

**MLE approach to the generalized Brody distribution**

As mentioned in previous sections, we have employed the generalized Brody distribution in the form which the parameters $\alpha$ and $\beta$ would assume as constant quantities. This is caused by troubles which occur in maximizing the Likelihood function contains Gamma functions, although we would display in the Figure 2, a closer corresponding are apparent between this definition and main distribution. The new distribution is

$$P(s) = b(1+q)(\alpha s^q + \beta s^{q+1})e^{-bs^{q+1}}, \qquad (I-1)$$

With multiplication of all P(s)'s, we can introduce likelihood function as

$$L(q,b,\alpha,\beta) = \prod_{i=1}^{n} b(1+q)\left(\alpha s_i^q + \beta s_i^{q+1}\right)e^{-bs_i^{q+1}} \qquad (I-2a)$$

Or

$$L(q,b,\alpha,\beta) = (b(1+q))^n \prod_{i=1}^{n}\left(\alpha s_i^q + \beta s_i^{q+1}\right)e^{-bs_i^{q+1}}, \qquad (I-2b)$$

We will use the logarithm of Eq.(I-2) to introduce the estimators of all variables as

$$\ln L(q,b,\alpha,\beta) = n\ln(b(1+q)) + \sum_{i=1}^{n}\ln(\alpha s_i^q + \beta s_i^{q+1}) - b\sum_{i=1}^{n} s_i^{q+1} \qquad (I-2c)$$

$$\frac{\partial \ln L(q,b,\alpha,\beta)}{\partial q} = 0 \Rightarrow f_1: \frac{n}{1+q} + \sum_{i=1}^{n}\ln s_i - b\sum_{i=1}^{n}\ln s_i\, s_i^{q+1} \qquad \text{for } q \qquad (I-3a)$$

$$\frac{\partial \ln L(q,b,\alpha,\beta)}{\partial b} = 0 \Rightarrow f_2: \frac{n}{b} - \sum_{i=1}^{n} s_i^{q+1} \qquad \text{for } b \qquad (I-3b)$$

$$\frac{\partial \ln L(q,b,\alpha,\beta)}{\partial \alpha} = 0 \Rightarrow f_3: \sum_{i=1}^{n}\frac{s_i^q}{\alpha s_i^q + \beta s_i^{q+1}} \qquad \text{for } \alpha \qquad (I-3c)$$

$$\frac{\partial \ln L(q,b,\alpha,\beta)}{\partial \beta} = 0 \Rightarrow f_4: \sum_{i=1}^{n}\frac{s_i^{q+1}}{\alpha s_i^q + \beta s_i^{q+1}} \qquad \text{for } \beta \qquad (I-3d)$$

We must take the derivates of all $f_i$ with related to all four variables to construct our Jacobian matrix for Newton-Raphson iteration method as[12-13,16]

$$\frac{\partial f_1}{\partial q} = -\frac{n}{(1+q)^2} - b\sum_{i=1}^{n}(\ln s_i)^2 s_i^{q+1} \qquad (I-4a)$$

$$\frac{\partial f_1}{\partial b} = -\sum_{i=1}^{n} \ln s_i \, s_i^{q+1} \qquad (I-4b)$$

$$\frac{\partial f_1}{\partial \alpha} = 0 \qquad (I-4c)$$

$$\frac{\partial f_1}{\partial \beta} = 0 \qquad (I-4d)$$

And similarly, for second estimator

$$\frac{\partial f_2}{\partial q} = -\sum_{i=1}^{n} s_i^{q+1} \ln s_i \qquad (I-5a)$$

$$\frac{\partial f_2}{\partial b} = -\frac{n}{b^2} \qquad (I-5b)$$

$$\frac{\partial f_2}{\partial \alpha} = 0 \qquad (I-5c)$$

$$\frac{\partial f_2}{\partial \beta} = 0 \qquad (I-5d)$$

And for third estimator

$$\frac{\partial f_3}{\partial q} = 0 \qquad (I-6a)$$

$$\frac{\partial f_3}{\partial b} = 0 \qquad (I-6b)$$

$$\frac{\partial f_3}{\partial \alpha} = -\sum_{i=1}^{n} \frac{s_i^{2q}}{\left(\alpha s_i^{q} + \beta s_i^{q+1}\right)^2} \qquad (I-6c)$$

$$\frac{\partial f_3}{\partial \beta} = -\sum_{i=1}^{n} \frac{s_i^{2q+1}}{\left(\alpha s_i^{q} + \beta s_i^{q+1}\right)^2} \qquad (I-6d)$$

And for fourth one, we have

$$\frac{\partial f_4}{\partial q} = 0 \qquad (I-7a)$$

$$\frac{\partial f_4}{\partial b} = 0 \qquad (I-7b)$$

$$\frac{\partial f_4}{\partial \alpha} = -\sum_{i=1}^{n} \frac{s_i^{2q+1}}{\left(\alpha s_i^q + \beta s_i^{q+1}\right)^2} \qquad (I-7c)$$

$$\frac{\partial f_4}{\partial \beta} = -\sum_{i=1}^{n} \frac{s_i^{2q+2}}{\left(\alpha s_i^q + \beta s_i^{q+1}\right)^2} \qquad (I-7d)$$

Now, we can apply Newton-Raphson iteration method as [12-13]

$$x_{new}^i = x_{old}^i - Df^{-1}(x_{old}^i) \qquad x^i: q, b, \alpha, \beta \qquad (I-8a)$$

$$\begin{bmatrix} q_{new} \\ b_{new} \\ \alpha_{new} \\ \beta_{new} \end{bmatrix} = \begin{bmatrix} q_{old} \\ b_{old} \\ \alpha_{old} \\ \beta_{old} \end{bmatrix} - Df^{-1}(q_{old}, b_{old}, \alpha_{old}, \beta_{old}) f(q_{old}, b_{old}, \alpha_{old}, \beta_{old}) \qquad (I-8b)$$

With applying these relations to our case, final results in order to evaluate our four parameters are obtained:

$$Denominator: \left[-\sum_{i=1}^{n} \ln s_i \, s_i^{q+1}\right]^2 \left[\sum_{i=1}^{n} \frac{s_i^{2q+1}}{\left(\alpha s_i^q + \beta s_i^{q+1}\right)^2}\right]^2 -$$

$$-\left[-\frac{n}{(1+q)^2} - b\sum_{i=1}^{n}(\ln s_i)^2 s_i^{q+1}\right]\left[-\frac{n}{b^2}\right]\left[\sum_{i=1}^{n} \frac{s_i^{2q+1}}{\left(\alpha s_i^q + \beta s_i^{q+1}\right)^2}\right]^2 -$$

$$-\left[-\sum_{i=1}^{n} \ln s_i \, s_i^{q+1}\right]^2 \left[-\sum_{i=1}^{n} \frac{s_i^{2q}}{\left(\alpha s_i^q + \beta s_i^{q+1}\right)^2}\right]\left[-\sum_{i=1}^{n} \frac{s_i^{2q+2}}{\left(\alpha s_i^q + \beta s_i^{q+1}\right)^2}\right] +$$

$$+\left[-\frac{n}{(1+q)^2} - b\sum_{i=1}^{n}(\ln s_i)^2 s_i^{q+1}\right]\left[-\frac{n}{b^2}\right]\left[-\sum_{i=1}^{n} \frac{s_i^{2q}}{\left(\alpha s_i^q + \beta s_i^{q+1}\right)^2}\right]\left[-\sum_{i=1}^{n} \frac{s_i^{2q+2}}{\left(\alpha s_i^q + \beta s_i^{q+1}\right)^2}\right] \qquad (I-9)$$

$$q_{new} = q_{old} - \{\frac{-\left[-\sum_{i=1}^{n}\frac{s_i^{2q+1}}{(\alpha s_i^q + \beta s_i^{q+1})^2}\right]^2 \left[-\frac{n}{b^2}\right]}{\text{Denominator}} +$$

$$+\frac{\left[-\sum_{i=1}^{n}\frac{s_i^{2q+2}}{(\alpha s_i^q + \beta s_i^{q+1})^2}\right]\left[-\sum_{i=1}^{n}\frac{s_i^{2q}}{(\alpha s_i^q + \beta s_i^{q+1})^2}\right]\left[-\frac{n}{b^2}\right]}{\text{Denominator}}\} \times$$

$$\times \left\{\frac{n}{1+q} + \sum_{i=1}^{n} \ln s_i - b \sum_{i=1}^{n} \ln s_i \, s_i^{q+1}\right\} +$$

$$+\{\frac{\left[-\sum_{i=1}^{n}\frac{s_i^{2q+1}}{(\alpha s_i^q + \beta s_i^{q+1})^2}\right]^2 \left[-\sum_{i=1}^{n} \ln s_i \, s_i^{q+1}\right]}{\text{Denominator}} -$$

$$-\frac{\left[-\sum_{i=1}^{n} \ln s_i \, s_i^{q+1}\right]\left[-\sum_{i=1}^{n}\frac{s_i^{2q+2}}{(\alpha s_i^q + \beta s_i^{q+1})^2}\right]\left[-\sum_{i=1}^{n}\frac{s_i^{2q}}{(\alpha s_i^q + \beta s_i^{q+1})^2}\right]}{\text{Denominator}}\} \times \left\{\frac{n}{b} - \sum_{i=1}^{n} s_i^{q+1}\right\} \quad (I-10)$$

$$b_{new} = b_{old} - \{\frac{\left[-\sum_{i=1}^{n}\frac{s_i^{2q+1}}{(\alpha s_i^q + \beta s_i^{q+1})^2}\right]^2 \left[-\sum_{i=1}^{n} \ln s_i \, s_i^{q+1}\right]}{\text{Denominator}} -$$

$$-\frac{\left[-\sum_{i=1}^{n}\frac{s_i^{2q+2}}{(\alpha s_i^q + \beta s_i^{q+1})^2}\right]\left[-\sum_{i=1}^{n}\frac{s_i^{2q}}{(\alpha s_i^q + \beta s_i^{q+1})^2}\right]\left[-\sum_{i=1}^{n} \ln s_i \, s_i^{q+1}\right]}{\text{Denominator}}\} \times$$

$$\times \left\{\frac{n}{1+q} + \sum_{i=1}^{n} \ln s_i - b \sum_{i=1}^{n} \ln s_i \, s_i^{q+1}\right\} +$$

$$+\{\frac{-\left[-\sum_{i=1}^{n}\frac{s_i^{2q+1}}{(\alpha s_i^q + \beta s_i^{q+1})^2}\right]^2 \left[-\frac{n}{(1+q)^2} - b \sum_{i=1}^{n} (\ln s_i)^2 s_i^{q+1}\right]}{\text{Denominator}} +$$

$$+\frac{\left[-\sum_{i=1}^{n}\frac{s_i^{2q+2}}{(\alpha s_i^q + \beta s_i^{q+1})^2}\right]\left[-\sum_{i=1}^{n}\frac{s_i^{2q}}{(\alpha s_i^q + \beta s_i^{q+1})^2}\right]\left[-\frac{n}{(1+q)^2} - b \sum_{i=1}^{n} (\ln s_i)^2 s_i^{q+1}\right]}{\text{Denominator}}\} \times$$

$$\times \left\{\frac{n}{b} - \sum_{i=1}^{n} s_i^{q+1}\right\} \quad (I-11)$$

$$\alpha_{new} = \alpha_{old} - \left\{ \frac{-\left[-\sum_{i=1}^{n} \ln s_i \, s_i^{q+1}\right]^2 \left[-\sum_{i=1}^{n} \frac{s_i^{2q+2}}{\left(\alpha s_i^q + \beta s_i^{q+1}\right)^2}\right]}{\text{Denominator}} + \right.$$

$$\left. + \frac{\left[-\frac{n}{(1+q)^2} - b \sum_{i=1}^{n} (\ln s_i)^2 s_i^{q+1}\right] \left[-\sum_{i=1}^{n} \frac{s_i^{2q+2}}{\left(\alpha s_i^q + \beta s_i^{q+1}\right)^2}\right] \left[-\frac{n}{b^2}\right]}{\text{Denominator}} \right\} \times \left\{ \sum_{i=1}^{n} \frac{s_i^q}{\alpha s_i^q + \beta s_i^{q+1}} \right\} +$$

$$+ \left\{ \frac{\left[-\sum_{i=1}^{n} \ln s_i \, s_i^{q+1}\right]^2 \left[-\sum_{i=1}^{n} \frac{s_i^{2q+1}}{\left(\alpha s_i^q + \beta s_i^{q+1}\right)^2}\right]}{\text{Denominator}} - \right.$$

$$\left. - \frac{\left[-\frac{n}{(1+q)^2} - b \sum_{i=1}^{n} (\ln s_i)^2 s_i^{q+1}\right] \left[-\frac{n}{b^2}\right] \left[-\sum_{i=1}^{n} \frac{s_i^{2q+1}}{\left(\alpha s_i^q + \beta s_i^{q+1}\right)^2}\right]}{\text{Denominator}} \right\} \times \left\{ \sum_{i=1}^{n} \frac{s_i^{q+1}}{\alpha s_i^q + \beta s_i^{q+1}} \right\} \quad (I-12)$$

$$\beta_{new} = \beta_{old} - \left\{ \frac{\left[-\sum_{i=1}^{n} \ln s_i \, s_i^{q+1}\right]^2 \left[-\sum_{i=1}^{n} \frac{s_i^{2q+1}}{\left(\alpha s_i^q + \beta s_i^{q+1}\right)^2}\right]}{\text{Denominator}} - \right.$$

$$\left. - \frac{\left[-\sum_{i=1}^{n} \frac{s_i^{2q+1}}{\left(\alpha s_i^q + \beta s_i^{q+1}\right)^2}\right] \left[-\frac{n}{(1+q)^2} - b \sum_{i=1}^{n} (\ln s_i)^2 s_i^{q+1}\right] \left[-\frac{n}{b^2}\right]}{\text{Denominator}} \right\} \times \left\{ \sum_{i=1}^{n} \frac{s_i^q}{\alpha s_i^q + \beta s_i^{q+1}} \right\} +$$

$$+ \left\{ \frac{-\left[-\sum_{i=1}^{n} \ln s_i \, s_i^{q+1}\right]^2 \left[-\sum_{i=1}^{n} \frac{s_i^{2q}}{\left(\alpha s_i^q + \beta s_i^{q+1}\right)^2}\right]}{\text{Denominator}} + \right.$$

$$\left. + \frac{\left[-\frac{n}{(1+q)^2} - b \sum_{i=1}^{n} (\ln s_i)^2 s_i^{q+1}\right] \left[-\frac{n}{b^2}\right] \left[-\sum_{i=1}^{n} \frac{s_i^{2q}}{\left(\alpha s_i^q + \beta s_i^{q+1}\right)^2}\right]}{\text{Denominator}} \right\} \times \left\{ \sum_{i=1}^{n} \frac{s_i^{q+1}}{\alpha s_i^q + \beta s_i^{q+1}} \right\} \quad (I-13)$$

To estimate with high accuracy by MLE technique, we have followed the prescription explained in Ref.[16], namely, the ML-based estimated parameters correspond to the converging values of iterations (I-10 to 13) where as an initial value, we have chosen the values of parameters obtained by LSF method. Now, if we consider the variations of A (the ratio of α to its exact value comes from Eq.(2.9)) and B (the ratio of β to it's exact value comes from Eq.(2.9)) in the iteration processes, an exact correspondence is yield. It means, our suggestion wouldn't apply any change to main definition, as have displayed in Figure2.

# Appendix (II)

# CRLB for new distribution

As have explained in Ref.[16], we must use the vector form of CRLB to describe the variations of related estimators. The Cramer-Rao Lower Bound inequality was defined [12-13]

$$cov_\theta(T(X)) \geq \frac{\partial \rho(\theta)}{\partial \theta^T}[F(\theta)]^{-1}\frac{\partial \rho^T(\theta)}{\partial \theta}, \qquad (II-1)$$

The CRLB for considered estimators will be as

$$\text{CRLB: } \frac{\partial \rho(\theta)}{\partial \theta^T}[F(\theta)]^{-1}\frac{\partial \rho^T(\theta)}{\partial \theta}\bigg|_{\text{with final values of } \alpha, \beta, b \text{ and } q \text{ obtained of MLE method}}, \qquad (II-2)$$

Now for our distribution, we have

$$\theta_1 \to q, \theta_2 \to b$$

And

$$\rho_1 \to \frac{1}{1+q} \Rightarrow \frac{\partial \rho_1}{\partial q} = \frac{-1}{(1+q)^2}, \quad \frac{\partial \rho_1}{\partial b} = 0 \quad \& \quad \rho_2 \to \frac{1}{b} \Rightarrow \frac{\partial \rho_2}{\partial q} = 0, \quad \frac{\partial \rho_2}{\partial q} = \frac{-1}{b^2}, \qquad (II-3)$$

On the other hand, for Fisher integral

$$F(\theta) = \begin{bmatrix} E\left[(X_q - \bar{X}_q)^2\right] & E[(X_q - \bar{X}_q)(X_b - \bar{X}_b)] \\ E[(X_q - \bar{X}_q)(X_b - \bar{X}_b)] & E[(X_b - \bar{X}_b)^2] \end{bmatrix}, \qquad (II-4)$$

Where

$$X_q = \frac{\partial \ln L(q,b)}{\partial q} = \frac{n}{1+q} + \sum_{i=1}^{n} \ln s_i - b\sum_{i=1}^{n} \ln s_i \, s_i^{q+1} \quad \& \quad \bar{X}_q = \frac{1}{n}\sum X_q \qquad (II-5)$$

$$X_b = \frac{\partial \ln L(q,b)}{\partial b} = \frac{n}{b} - \sum_{i=1}^{n} s_i^{q+1} \quad \& \quad \bar{X}_b = \frac{1}{n}\sum X_b \qquad (II-6)$$

Which can combine to final form

$$CRLB: \begin{pmatrix} \frac{\partial \rho_1}{\partial q} & \frac{\partial \rho_1}{\partial b} \\ \frac{\partial \rho_2}{\partial q} & \frac{\partial \rho_2}{\partial q} \end{pmatrix} \begin{bmatrix} E\left[(X_q - \bar{X}_q)^2\right] & E[(X_q - \bar{X}_q)(X_b - \bar{X}_b)] \\ E[(X_q - \bar{X}_q)(X_b - \bar{X}_b)] & E[(X_b - \bar{X}_b)^2] \end{bmatrix}^{-1} \begin{pmatrix} \frac{\partial \rho_1}{\partial q} & \frac{\partial \rho_2}{\partial q} \\ \frac{\partial \rho_1}{\partial b} & \frac{\partial \rho_2}{\partial q} \end{pmatrix} \qquad (II-7)$$

Or

$$CRLB: \frac{\left[E\left[(X_q - \bar{X}_q)^2\right] E[(X_b - \bar{X}_b)^2] - (E[(X_q - \bar{X}_q)(X_b - \bar{X}_b)])^2\right]}{b^4(1+q)^4(E\left[(X_q - \bar{X}_q)^2\right] E[(X_b - \bar{X}_b)^2] - (E[(X_q - \bar{X}_q)(X_b - \bar{X}_b)])^2)} \qquad (II-8)$$

# Figure caption

**Figure1.** NNSDs for graphene with different magnetic fields. Solid, dashed and dotted lines represent, GUE, Poisson and GOE limits respectively.

**Figure2.** (color online).The variation of our proposed constants for new distribution in different iteration stages which verify our aim that any change wouldn't occur in compare to the main distribution .The left one represented for $\alpha$ which horizontal axis represent number of iteration and vertical one represent A (the ratio of $\alpha$ to its exact value comes from Eq.(2.9)) and the right one display variation of $\beta$ which the horizontal axis represent number of iteration and vertical one represent B (the ratio of $\beta$ to it's exact value comes from Eq.(2.9)).

Figure1.

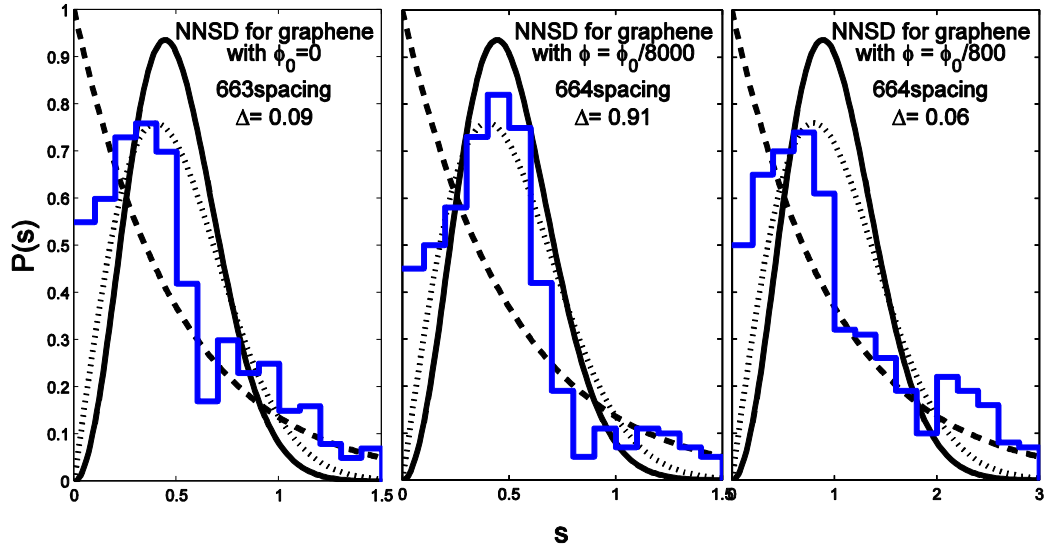

Figure2.

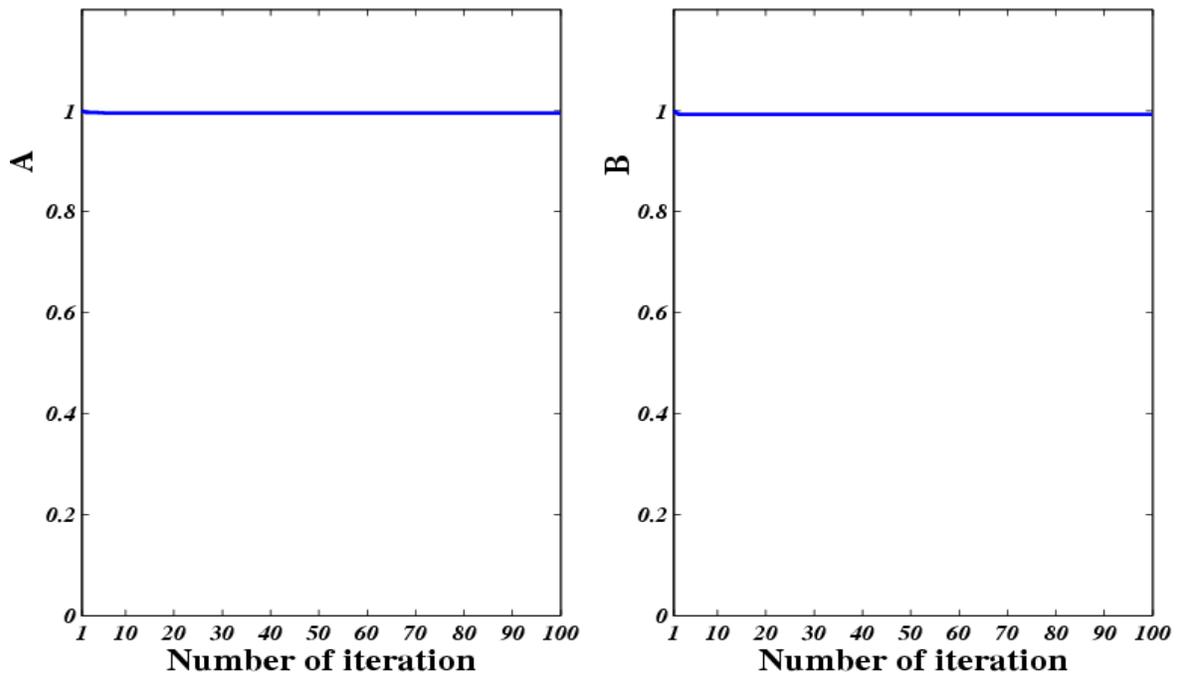